# Role of Statistical tests in Estimation of the Security of a New Encryption Algorithm

*A.V.N. Krishna,*
*Professor, Computer Science Department, Indur Institute of Engineering & Technology, Siddipet, Andhra Pradesh, India.*
*Email: hari_avn@Rediffmail.com*
*A.Vinaya Babu,*
*Director, Admissions, J.N.T.U, Hyderabad.*

**Abstract**

Encryption study basically deals with three levels of algorithms. The first algorithm deals with encryption mechanism, second deals with decryption Mechanism and the third discusses about the generation of keys and sub keys used in the encryption study. In the given study, a new algorithm is discussed which generates a random sequence which is used as sub key for encryption process. The algorithm executes a series of steps and generates a sequence. This sequence is being used as sub key to be mapped to plain text to generate cipher text. The strength of the encryption & Decryption process depends on the strength of sequence generated against crypto analysis. In this part of work some statistical tests like Uniformity tests, Universal tests & Repetition tests are tried on the sequence generated to test the strength of it.

## 1. Introduction

A crypto system [1-5] is an algorithm, plus all possible plain texts, cipher texts and keys. There are two general types of key based algorithms: symmetric and public key. With **symmetric-key encryption,** the encryption key can be calculated from the decryption key and vice versa. With most symmetric algorithms, the same key is used for both encryption and decryption.

Implementations of symmetric-key encryption can be highly efficient, so that users do not experience any significant time delay as a result of the encryption and decryption. Symmetric-key encryption also provides a degree of authentication, since information encrypted with one symmetric key cannot be decrypted with any other symmetric key. Thus, as long as the symmetric key is kept secret by the two parties using it to encrypt communications, each party can be sure that it is communicating with the other as long as the decrypted messages continue to make sense.

Symmetric-key encryption is effective only if the symmetric key is kept secret by the two parties involved. If anyone else discovers the key, it affects both confidentiality and authentication. A person with an unauthorized symmetric key not only can decrypt messages sent with that key, but can encrypt new messages and send them as if they came from one of the two parties who were originally using the key.

## 2. Algorithm [6-11]

The new algorithm has the following features

1. A set of poly alphabetic substitution rule is used.
2. A new block cipher is developed.
3. A Random matrix is being used as key.
4. Generated sequence being used as sub key.
5. Coding method.

The steps that are involved in the proposed algorithm.

1. A sequence of n values say n=0-26 is considered.
2. This sequence is represented in ternary format.
3. Let this be "**B**".
4. 1 is subtracted from all the values of ternary vector.
5. A random matrix is used as a key. Let it be A**.**
6. The modified ternary vector is multiplied with the matrix key.
7. A sign function is applied on the product of ternary vector & matrix key.
8. 1 is added to all values of step 7.
9. A sequence is generated which is used as sub key
10. Equality of values of the sequence generated by step 9 are considered to develop a random sequence.

**Example**

**Step1:**

Consider the sequence for n= 0 to 26 values.

**Step2:**

Convert the sequence to ternary form of a 3 digit number

i.e. 0 ------- 000
1-------- 001
2-------- 002
.
.
.
26-------- 222

**Step3**:

Represent above ternary form in 27x3 matrix

$$R = \begin{bmatrix} 0 & 0 & 0 \\ 0 & 0 & 1 \\ 0 & 0 & 2 \\ 0 & 1 & 0 \\ 0 & 1 & 1 \\ 0 & 1 & 2 \\ 0 & 2 & 0 \\ 0 & 2 & 1 \\ 0 & 2 & 2 \\ 1 & 0 & 0 \\ 1 & 0 & 1 \\ 1 & 0 & 2 \\ 1 & 1 & 0 \\ 1 & 1 & 1 \\ 1 & 1 & 2 \\ 1 & 2 & 0 \\ 1 & 2 & 1 \\ 1 & 2 & 2 \\ 2 & 0 & 0 \\ 2 & 0 & 1 \\ 2 & 0 & 2 \\ 2 & 1 & 0 \\ 2 & 1 & 1 \\ 2 & 1 & 2 \\ 2 & 2 & 0 \\ 2 & 2 & 1 \\ 2 & 2 & 2 \end{bmatrix}$$

**Step 4:**

Subtract 1 from each element of the above matrix and the resulting matrix R is

$$R = \begin{bmatrix} -1 & -1 & -1 \\ -1 & -1 & 0 \\ -1 & -1 & 1 \\ -1 & 0 & -1 \\ -1 & 0 & 0 \\ -1 & 0 & 1 \\ -1 & 1 & -1 \\ -1 & 1 & 0 \\ -1 & 1 & 1 \\ 0 & -1 & -1 \\ 0 & -1 & 0 \\ 0 & -1 & 1 \\ 0 & 0 & -1 \\ 0 & 0 & 0 \\ 0 & 0 & 1 \\ 0 & 1 & -1 \\ 0 & 1 & 0 \\ 0 & 1 & 1 \\ 1 & -1 & -1 \\ 1 & -1 & 0 \\ 1 & -1 & 1 \\ 1 & 0 & -1 \\ 1 & 0 & 0 \\ 1 & 0 & 1 \\ 1 & 1 & -1 \\ 1 & 1 & 0 \\ 1 & 1 & 1 \end{bmatrix}$$

**Step5:**

Consider a random matrix

$$A = \begin{bmatrix} 2 & 5 & -6 \\ 3 & 1 & 3 \\ 4 & -2 & -3 \end{bmatrix}$$

**Step6:**

R= R X A

$$R = \begin{bmatrix} -1 & -7 & 1 \\ -7 & -4 & -2 \\ -13 & -1 & -5 \\ 4 & -6 & -1 \\ -2 & -31 & -4 \\ -8 & 0 & -7 \\ 9 & -5 & -3 \\ 3 & -2 & -6 \\ -3 & 1 & -9 \\ 1 & -4 & 5 \\ -5 & -1 & 2 \\ -11 & 2 & -1 \\ 6 & -3 & 3 \\ 0 & 0 & 0 \\ -6 & 3 & -3 \\ 11 & -2 & 1 \\ 5 & 1 & -2 \\ -1 & 4 & -5 \\ 3 & -1 & 9 \\ -3 & 2 & 6 \\ -9 & 5 & 3 \\ 8 & 0 & 7 \\ 2 & 3 & 4 \\ -4 & 6 & 1 \\ 13 & 1 & 5 \\ 7 & 4 & 2 \\ 1 & 7 & -1 \end{bmatrix}$$

**Step7:**

Convert all positive values to 1, negative values to -1 and zero to 0 of the resulting matrix of step 6.

$$R = \begin{bmatrix} -1 & -1 & 1 \\ -1 & -1 & -1 \\ -1 & -1 & -1 \\ 1 & -1 & -1 \\ -1 & -1 & -1 \\ -1 & 0 & -1 \\ 1 & -1 & -1 \\ 1 & -1 & -1 \\ -1 & 1 & -1 \\ 1 & -1 & 1 \\ -1 & -1 & 1 \\ -1 & 1 & -1 \\ 1 & -1 & 1 \\ 0 & 0 & 0 \\ -1 & 1 & -1 \\ 1 & -1 & 1 \\ 1 & 1 & -1 \\ -1 & 1 & -1 \\ 1 & -1 & 1 \\ -1 & 1 & 1 \\ -1 & 1 & 1 \\ 1 & 0 & 1 \\ 1 & 1 & 1 \\ -1 & 1 & 1 \\ 1 & 1 & 1 \\ 1 & 1 & 1 \\ 1 & 1 & -1 \end{bmatrix}$$

**Step8:** Add 1 to each element of the matrix R

$$R = \begin{bmatrix} 0 & 0 & 2 \\ 0 & 0 & 0 \\ 0 & 0 & 0 \\ 2 & 0 & 0 \\ 0 & 0 & 0 \\ 0 & 1 & 0 \\ 2 & 0 & 0 \\ 2 & 0 & 0 \\ 0 & 2 & 0 \\ 2 & 0 & 2 \\ 0 & 0 & 2 \\ 0 & 2 & 0 \\ 2 & 0 & 2 \\ 1 & 1 & 1 \\ 0 & 2 & 0 \\ 2 & 0 & 2 \\ 2 & 2 & 0 \\ 0 & 2 & 0 \\ 2 & 0 & 2 \\ 0 & 2 & 2 \\ 0 & 2 & 2 \\ 2 & 1 & 2 \\ 2 & 2 & 2 \\ 0 & 2 & 2 \\ 2 & 2 & 2 \\ 2 & 2 & 2 \\ 2 & 2 & 0 \end{bmatrix}$$

**Step9:**

Convert each row of the matrix R to decimal form to generate sequence.
ie 0 0 2 will form $0*3^{2}+0*3^{1}+2*3^{0} = 2$

The sequence formed is = 2 0 0 18 0 3 18 18 6 20 2 6 20 13 6 20 24 6 20 8 8 23 26 8 26 26 24.

**Generating Sub keys from sequence generated:**

1. n [27]= 0 1 2 3 4 5……………………………………………...26

2. r [27] = 2 0 0 18 0 3 18 18 6 20 2 6 20 13 6 20 24 6 20 8 8 23 26 8 26 26 24.

3. Read n[0]=0. Store the values of n[0],r[0] in a basin. ie b(0)=(0,2).

4.1. n[0] ie. ‘0’ is compared with r[27] values. There is a match at r[1],r[2] & r[4]. Neglect already visited elements. Thus b(0)=[0,1,2,4).

4.2 Step 4.1 is repeated with other elements of basin ie. 1, 2 & 4 values. For elements 1& 4, there is no match of values in r[27]. For element 2, there is a match at r[10]. Thus the basin b(0) is updated to (0, 2, 1, 4 &10).

5. The procedure is repeated for the next element of the sequence of step 1 which is not visited earlier. The random sequence formed is

(0,2,1,4,10,3,18,5,6,7,8,11,14,17,19,20,23,9,12,15,21,13,24,26,16,22,25)

This can be used as sub keys in encryption decryption process.

## 4. Case Studies:

Different cases using different keys have been tried to test the strength of algorithm against crypto analysis.

**Case 1:**

Considering the key A of size (3*3)

$$A = \text{key} \begin{matrix} 2 & 5 & -6 \\ 3 & 1 & 3 \\ 4 & -2 & -3 \end{matrix}$$

Sequence generated from the proposed model

r=2 0 0 18 0 3 18 18 6 20 2 6 20 13 6 20 24 6 20

r= 8 8 23 26 8 26 26 24.

Considering the equality of values, the random sequence that can be formed using this sequences are

(0,2,1,4,10,3,18,5,6,7,8,11,14,17,19,20,23,9,12,15,21,13,24,26,16,22,25)

**Case 2**: By increasing the values of key by 1 at each row

$$A = \text{key} = \begin{matrix} 3 & 5 & -6 \\ 4 & 1 & 3 \\ 5 & -2 & -3 \end{matrix}$$

Sequence generated

r= 1 0 0 18 0 0 18 18 3 20 2 6 20 13 6 20 24 6 23 8 8 26 26 8 26 26 25.

The random sequence generated from this sequence is

0,1,2,4,5,10,3,18,8,6,7,19,20,23,11,14,17,9,12,15,13,24,6,16,21,22,25.

**Case 3:**

By decreasing the key values by 1 at each row.

$$A = key = \begin{matrix} 1 & 5 & -6 \\ 2 & 1 & 3 \\ 2 & -2 & -3 \end{matrix}$$

Sequence generated by the model.

r= 11,1,3,20,0,6,18,18,6,20,2,6,20,13,6,20,24,6,20,8,8,20,26,6,23,25,15.

The random sequence generated from this sequence is

0,11,4,1,2,3,10,6,18,5,8,14,17,23,7,19,20,24,9,12,15,21,16,26,22,13,25

**Case 4:**

By increasing the key size to [4*4], we can increase the number of values of the sequence, which increases the strength of the algorithm.

$$A = key = \begin{matrix} 1 & 5 & -6 & 1 \\ 2 & 1 & 3 & 2 \\ 3 & -2 & -3 & 3 \\ 4 & 2 & 4 & 4 \end{matrix}$$

Sequence generated for n=0 to 80 values represented to the base 3.

r=0,33,60,0,0,6,0,9,20,54,57,60,0,0,40,0,19,20,54,54,60,54,54,74,9,20,20,33,60,60,0,6,26,9,20,26,57,60,61,0,40,80,19,20,23,54,60,71,54,74,80,20,20,47,60,60,71,6,26,26,20,26,26,60,61,80,40,80,80,20,23,26,60,71,80,74,80,80,20,47,80.

The random sequence formed by considering similar values,

0,3,4,6,12,13,15,30,39,5,31,57,10,36,9,54,7,24,33,18,19,21,22,45,48,1,27,16,42,20,60,8,17,25,26,34,43,51,52,69,78,2,11,28,29,37,46,54,55,63,72,32,35, 58,59,61,62,71,9,18,19,21,22,45,48,38,64,47,56,73,7,24,33,16,42,53,79,1,27,23,74,44,70,49,75,40,14, 66,80,41,50,65,67,68,76,77.

**Case 5:**

$$A = \quad key = \begin{matrix} 2 & 5 & -6 & 1 \\ 3 & 1 & 3 & 2 \\ 4 & -2 & -3 & 3 \\ 5 & 2 & 4 & 4 \end{matrix}$$

Sequence generated n=0 to 80 values represented to the base 3.

r=0,6,33,0,0,6,0,0,20,54,54,60,0,0,0,0,9.20,54,54,57,54,54,65,0,20,20,33,60,60,0,6,26,9,20,26,57,60,61,0,40,80,19,20,23,24,60,71,54,74,80,20,20,47,60,60,80,15,26,26,23,26,26,60,71,80,80,80,80,20,26,26,26,16,80,80,74,80,80,47,74,80.

The random sequence formed from the generated sequence

0,3,4,6,7,12,13,14,15,24,30,39,1,5,31,57,45,20,36,8,17,25,26,34,43,44,51,52,69,32,35,58,59,61,62,70,71,38,47,64,53,78,9,54,16,33,10,18,19,21,27,48,72,2,42,23,65,60,11,28,29,37,46,54,55,63,9,10,18,19,21,22,48,16,33,42,72,2,27,40.

**Case 6:**

$$A = key = \begin{matrix} 3 & 5 & -6 & 1 \\ 4 & 1 & 3 & 2 \\ 5 & -2 & -3 & 3 \\ 6 & 2 & 4 & 4 \end{matrix}$$

Sequence generated for n=0 to 80 values being represented to the base 3.

r=0,3,6,0,0,3,0,0,10,54,54,60,0,0,0,0,0,20,54,54,54,54,54,55,0,10,20,33,60,60,0,6,26,9,20,26,57,60,61,0,40,80,19,20,23,54,60,71,74,54,80,20,20,47,60,70,80,25,26,26,26,26,26,60,80,80,80,80,80,20,26,70,80,80,77,80,80,74,77,80.

The random sequence formed from the generated sequence is

0,3,4,6,7,12,13,14,15,16,22,28,37,79,80,1,5,2,29,39,48,54,62,63,64,65,66,67,71,72,74,75,78,9,10,18,9,20,43,47,45,46,76,31,8,23,40,17,24,32,41,49,50,68,51,42,38,25,33,55,21,26,60,30,56,57,58,59,69,11,27,35,44,52,61,34,36,70,53,77,73.

**5. Role of Statistical Tests on Sequence Generated:**

Uniform test tries to study the uniformity of a sequence generated. From the sequence generated, the first two consecutive points are considered as coordinates of a graph. The process is repeated for all the values of the sequence to generate the graph. For example, if a,b,c,d……..n are the values of sequence generated, then the coordinates of the graph are (a,b),(c,d)……….(n-1,n). If they are dependent they will form into lines or they will form into a plane. In this study, different cases are tried for both the algorithms to study the non uniformity distribution of values in the sequence.
One more test that can test the randomness of the sequence generated is Chi Square test. In this test the sum of $X^2$ is calculated as

$X^2$ = € (original value-expected value)$^2$/Expected value. This $X^2$ is being used to calculate the probability of occurrence of the values in the random sequence generated.

Universal tests: It tries to see whether the data can be compressed. Since in the given sequence, the repetition of values is minimum, we can say that the developed sequences are relatively free from this test.

Repetition test: This test studies that each character of the sequence is not identical to the earlier character. In the sequence generated by the algorithm, the values are random and unique in nature, the probability that the values get repeated is 1/27. Thus we can say that the sequence generated by the algorithm is relatively free from Uniformity test, Universal test & Repetition tests.

The analysis is being done on sequences generated by the algorithm which shows the non uniformity in the values of sequence generated which ultimately provides Maximum strength to the algorithm. The analysis is being represented as graphs for the discussed algorithm (see Fig.1).

When Chi-Square test is applied on the sequence generated in the different case studies, it is observed that the value of $X^2$ formed is around 160 with a probability of occurrence of values at a very low kevel of 0.06. This shows that the sequence generated in the different cases is highly random in nature which provides sufficient strength to the encryption algorithm against crypto analysis.

**6. Conclusions**

Using the developed algorithm, a sequence is generated which is used as sub key. Different statistical tests are applied on the sequence generated to test the validity of it. It is concluded that the developed algorithm is generating a random sequence which can be used as sub key which has sufficient strength against crypto analysis.

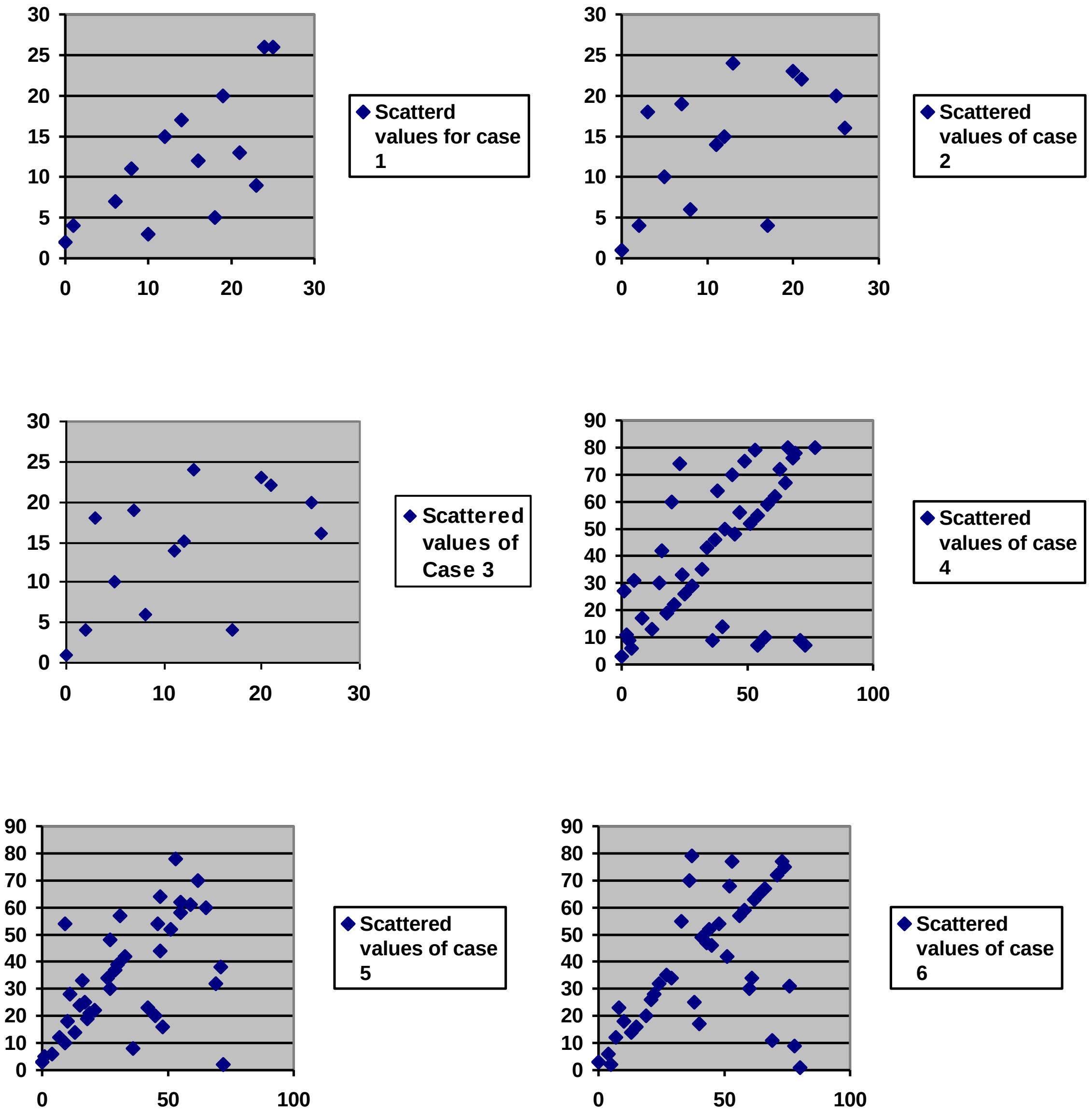

Figure 1: Sequential Data Representation of Algorithm